\documentclass[prl,onecolumn,10pt]{revtex4}
\usepackage{graphicx}
\usepackage{amsmath}
\usepackage{epsfig}
\usepackage{setspace}
\usepackage{anysize}
\begin{document}

\title{Interferometric weak value deflections: quantum and classical treatments}

\author{John C. Howell, David J. Starling, P. Ben Dixon, Praveen K. Vudyasetu, and Andrew N. Jordan}
\affiliation{Department of Physics and Astronomy, University of
Rochester, Rochester, NY 14627, USA }

\begin{abstract}
We derive the weak value deflection given in a paper by Dixon \textit{et al.}\ \cite{Dixon09} both quantum mechanically and classically. This paper is meant to cover some of the mathematical details omitted in that paper owing to space constraints.
\end{abstract}

\maketitle

Weak values \cite{Aharanov88} have presented or inspired intriguing possibilities for precision measurement.  A recent example is by Hosten and Kwiat \cite{Hosten08}, where they were able to amplify the deflections arising from the spin Hall effect of light.  The light fields used in their experiment as well as in \cite{Dixon09} were coherent quasi-classical fields and no apparent quantum mechanical system was employed in the experiments.  The classical behavior of these weak value inspired deflection experiments has been known for some time \cite{Ritchie91}.  Shortly after the Hosten and Kwiat paper, Aiello and Woerdman \cite{Aiello08} published the classical description to allow greater accessibility to the metrology community. Here, we derive both a quantum weak value amplification for a Sagnac interferometer \cite{Dixon09} along with its classical counterpart under the corresponding limits using the well developed understanding in classical interferometry.

Now consider the interferometric weak value experiment in \cite{Dixon09}.  We point out that all two dimensional quantum systems are isomorphic to spin-1/2 particles.  In the Hosten-Kwiat experiment, the two dimensional system was the transverse polarization states of the light.  For the weak value description in this paper, we use the which-path states of a photon in a Sagnac interferometer as the two-state system.   We first derive the quantum mechanical weak value description for a single photon and proceed to derive the classical field description.

For the quantum mechanical derivation we present the theory for a single photon in a Sagnac interferometer, the photon's which way variable (system) is coupled to its the transverse momentum (meter).  The system eigenstates are $\{\vert a_i\rangle\}$ and the meter eigenstates are $\{\vert k_x\rangle\}$. The pre-selected total state of the photon is the tensor product of the system and meter states, written as

\begin{equation}
\vert \Psi\rangle=\int dk_x\psi(k_x)a^\dagger_{k_x}\vert 0\rangle \vert\varphi_1\rangle,
\end{equation}
where $\vert\varphi_1\rangle=\sum_i c_i\vert a_i\rangle$, $\psi(k_x)$ is the transverse wavefunction. We will assume that this is a Gaussian in order to obtain an analytic solution.  Finally, $a^\dagger_{k_x}$ is the creation operator for the photon in the transverse mode.

We now describe the weak measurement procedure. First, the photon undergoes a small unitary evolution, which couples one of the propagation directions in the interferometer to one momentum and the other direction to another momentum.  In essence, the momentum shift, upon detection, gives a small amount of which-path information about the path of the photon. The unitary evolution is given by $U=e^{-i k \hat{A}x} \approx 1-i k \hat{A}x$, where $\hat{A}$ is the which-path observable with eigenvalues $\hat{A}\vert a_i\rangle=a_i\vert a_i\rangle$.  Measuring this small momentum shift constitutes the weak measurement.  In this scheme the weak measurement and post selection measurement happen simultaneously at the output port of the beamsplitter (measuring the transverse momentum, and postselecting on the output port)

For simplicity, the calculation here will assume a collimated beam with no divergence.  For this approach, a mirror imparts a weak transverse momentum shift $k$, in opposite directions relative to the optical axis at the exit face of the beam splitter.  As noted earlier, this deflection gives partial information about which way the photon went in the interferometer.  However, the momentum imparted to the photon gives a transverse shift to the photon in the detection plane.  For very small momentum shifts and short distances, the deflection is very small compared to the transverse diameter of the beam and thus the eigenstates are only weakly discriminated.  After passing through the beamsplitter, post-selection on the state  $\vert \varphi_2\rangle$, which is nearly orthogonal to the input state, is applied to the photon.  This yields a post-selected meter state

\begin{equation}
\langle \varphi_2\vert U \vert \Psi\rangle \approx \int dk_x\psi(k_x)a^\dagger_{k_x}\vert 0\rangle \langle \varphi_2 \vert\varphi_1\rangle-i\int dk_x  \psi(k_x) k x a^\dagger_{k_x}\vert 0\rangle \langle \varphi_2\vert \hat{A} \vert\varphi_1\rangle.
\end{equation}

As can be seen, if the pre- and the post-selected system state are nearly orthogonal, the probability for the photon to pass through through the post-selecting device (e.g., polarizer or beam splitter) is small.  However, for the photons that do pass through, we must renormalize the single photon meter state.  We define the renormalized state as

\begin{equation}
\vert \Psi'\rangle=\int dk_x\psi(k_x)a^\dagger_{k_x}\vert 0\rangle -i\int dk_x  \psi(k_x) k x a^\dagger_{k_x}\vert 0\rangle \frac{ \langle \varphi_2\vert \hat{A}
\vert\varphi_1\rangle}{\langle \varphi_2 \vert\varphi_1\rangle},\label{exponentiated}
\end{equation}
where the term $\frac{ \langle \varphi_2\vert \hat{A} \vert\varphi_1\rangle}{\langle \varphi_2 \vert\varphi_1\rangle}$ is the standard weak value term.   A quick example shows why this term is imaginary.  Suppose the pre-selected spin state is given by $\vert \varphi_1\rangle=\frac{1}{\sqrt{2}}(e^{-i\frac{\phi}{2}}\vert +\rangle +e^{i\frac{\phi}{2}}\vert -\rangle)$ and the post-selected state is $\vert \varphi_2\rangle=\frac{1}{\sqrt{2}}(\vert +\rangle -\vert -\rangle)$.  We then see that $\langle \varphi_2 \vert\varphi_1\rangle=i \sin(\frac{\phi}{2})$. If the observable $\hat{A}$ is the Pauli operator $\sigma_z$ we find $ \langle \varphi_2\vert \hat{A} \vert\varphi_1\rangle=\cos(\phi/2)$. Thus $\frac{ \langle \varphi_2\vert \hat{A} \vert\varphi_1\rangle}{\langle \varphi_2 \vert\varphi_1\rangle}$ is purely imaginary. Noting this, we let $A_w=\left \vert \frac{ \langle \varphi_2\vert \hat{A} \vert\varphi_1\rangle}{\langle \varphi_2 \vert\varphi_1\rangle}\right \vert$. In this example, small $\phi$ produces a large $A_w$. We will see that this corresponds to a standard weak value enhancement.

As long as the second term on the right hand side is much smaller than the first of Eq. (\ref{exponentiated}), we can reexponentiate to obtain
\begin{equation}
\vert \Psi'\rangle=\int dk_x\psi(k_x)e^{-k A_w x}a^\dagger_{k_x}\vert 0\rangle.
\end{equation}

To obtain the probability amplitude distribution in the transverse plane, we define a positive frequency field operator
\begin{equation}
E^+(x)=\int dk_x E_0 e^{-i k_x x}a_{k_x}.
\end{equation}
Incorporating this result and using the commutation relation $[a_{k_x'},a^\dagger_{k_x}]=\delta_{k_x',k_x}$, the statevector becomes
\begin{equation}
\langle 0 \vert E^+(x)\vert \Psi'\rangle=E_0 \int dk_xe^{-i k_x x}\psi(k_x)e^{-k A_w x}.
\end{equation}
From this point, we will not worry about the normalization of the state and use the Gaussian wavefunction.  Using the fact that $A_w\approx 2/\phi$ for small $\phi$, we find

\begin{equation}
\langle 0 \vert E^+(x)\vert \Psi'\rangle\propto e^\frac{-2 k x}{\phi}\int dk_x e^{-i k_x x}e^{- k_x^2 \sigma^2} = \exp\left[-\frac{x^2}{4\sigma^2}-\frac{2 k x}{\phi}\right],
\end{equation}
where $\sigma$ is the Gaussian beam radius.  After completing the square,
\begin{equation}
\langle 0 \vert E^+(x)\vert \Psi'\rangle \propto \exp\left[-\frac{1}{4\sigma^2}\left(x+\frac{4 k \sigma^2}{\phi}\right)^2\right].
\end{equation}
One can see that, at the detector, there will be a transverse position shift of beam given by $d_w=\frac{4 k \sigma^2}{\phi}$, where we denote $d_w$ as the weak value transverse deflection.

We now derive the same result classically using standard wave optics.  This can be done by denoting the transverse two-port input field of the interferometer as
\begin{equation}
E_{in}=\left(
         \begin{array}{c}
           E_0e^{-x^2/4\sigma^2} \\
           0 \\
         \end{array}
       \right),
\end{equation}
where the second position in the column vector denotes the input port with no electric field.  The field then passes through a 50/50 beamsplitter with a matrix representation
\begin{equation}
B=\frac{1}{\sqrt{2}}\left(
                      \begin{array}{cc}
                        1 & i \\
                        i & 1 \\
                      \end{array}
                    \right).
\end{equation}
We now define a matrix which gives both an opposite momentum shift $k$ and a relative phase between the two paths
\begin{equation}
M=\left(
\begin{array}{cc}
       e^{i(-k x+\phi/2)} & 0 \\
       0 &  e^{-i(-k x+\phi/2)}\\
\end{array}
\right).
\end{equation}
We want to determine the field at the ``dark" output port (i.e.\ the port with the lowest intensity of light coming out of it) of the interferometer.  The evolution of the light is represented by the matrix combination
\begin{equation}
E_{out}=(BMB) E_{in}.
\end{equation}

The output field at the dark port is renormalized by noting that the detector only measures the total flux falling on it to determine the deflection.  For small $k$, the measured output signal will be of the form
\begin{equation}
E_{out}^d=\frac{\sin(-k x+\phi/2)}{\sin(\phi/2)}\exp[-x^2/4\sigma^2],
\end{equation}
where the superscript $d$ denotes the renormalized (by $\sin(\phi/2)$) dark port of the interferometer.  For small angles, we obtain $(1-\frac{2 k x}{\phi})\exp[-x^2/4\sigma^2]$, which we reexponentiate, complete the square and obtain
\begin{equation}
E_{out}'\propto \exp\left[-\frac{1}{2\sigma^2}\left(x+\frac{4 k \sigma^2}{\phi}\right)^2\right].
\end{equation}
We see that we obtain the same result as the quantum mechanical weak value treatment.

To consider the case of a diverging beam we insert a negative focal length lens before the interferometer and use standard Fourier optics methods in the paraxial approximation outlined in Goodman \cite{Goodman}. In the case of the quantum treatment, phase factors and Fourier transforms are applied to the quantum state $\Psi(x)$ or $\Psi(k_x)$ by convention. Similarly, in the classical treatment, they are applied to the electric field $E$.

Passing through the lens, the wavefunction (electric field) acquires a multiplicative phase factor $\exp(ik_0 x^2/(2s_i))$, where $k_0$ is the wavenumber of the light and $s_i$ is the image distance behind the lens, resulting in a spreading beam. Propagation effects are accounted for by Fourier transforming the state (field) at the lens, and applying a multiplicative phase factor $\exp(-ip^2 l_{lm}/(2k_0))$ to the momentum-space wavefunction (field), where $l_{lm}$ is the distance from lens to mirror. The effect of the oscillating mirror is to shift the state (field) by a very small transverse momentum $k$, $\Phi(p) \rightarrow \Phi(p \pm k)$ ($E(k_x) \rightarrow E(k_x \pm k)$), where the direction of the shift depends on which path the photon takes in the interferometer. Propagation from mirror to detector results in a final multiplicative phase factor $\exp(-ip^2 l_{md}/(2k_0))$ on the momentum-space wavefunction (field), with $l_{md}$ being the distance from mirror to detector. The individual amplitudes in both arms are given by
\begin{equation}
\Psi_{1,2}(x) \propto \exp{\left[\frac{-ik_0 x^2 \pm 2ilkx}{2(l + l_{md})}\right]},
\end{equation}
up to normalization, where $l = l_{lm} - a^2 s_i /(a^2 + is_i /(2k_0))$ and $a$ is the beam radius at the lens. These amplitudes (fields) now interfere with a relative phase $\phi$, and the position of the beam is monitored with a quad detector at the dark port. Because the relative momentum shift $k$ given by the movable mirror is so small, the post-selection probability is given by the overlap of pre- and post-selected states, $P_{ps} = \sin^2(\phi/2) \approx \phi^2/4$ for $\phi \ll 1$ as before. Assuming the diffractive effects are small, so that the wavelength $\lambda \ll 2\pi a^2/s_i$, we find that the beam deflection is given by
\begin{equation}
d'_w=\frac{4 k a^2}{\phi} \frac{l_{im}(l_{im} + l_{md})}{s_i^2},
\end{equation}
where $l_{im}$ the length from the image to the mirror.

We have derived the weak value deflection measurement results in the paper by Dixon \textit{et al.} \cite{Dixon09} using both classical and quantum methods. The results for a diverging beam using classical Fourier techniques with quantum wave functions were shown in more detail.

This work was supported by DARPA DSO Slow Light, and a DOD PECASE award.


\begin{references}

\bibitem{Dixon09} P.B. Dixon, D.J. Starling, A.N. Jordan, and J.C. Howell, Phys. Rev. Lett. {\bf 102} 173601 (2009)

\bibitem{Aharanov88} Y. Aharanov, D.Z. Albert, and L. Vaidman, Phys. Rev. Lett. {\bf 60}, 1351 (1988)

\bibitem{Hosten08} O. Hosten, and P. Kwiat, Science {\bf 319} 787 (2008)

\bibitem{Aiello08} A. Aiello, and J.P. Woerdman, Opt. Lett. {\bf 33}, 1437 (2008)

\bibitem{Ritchie91} N.W.M. Ritchie, J.G. Story, and R.G. Hulet, Phys. Rev. Lett. {\bf 60}, 1351 (1988)

\bibitem{Goodman} J.W. Goodman, Introduction to Fourier Optics, McGraw-Hill, 1968

\end{references}
\end{document}